%% file: main.tex
\documentclass[letterpaper, 10 pt, conference]{ieeeconf}
\usepackage{cite}
\usepackage{comment}
\usepackage{makecell}
\usepackage{xcolor}
\usepackage{tikz}
\usepackage[normalem]{ulem}
\usepackage{cntsys}

\usepackage{subcaption}

\RequirePackage{amsthm}
\newtheorem{theorem}{Theorem}
\newtheorem{problem}{Problem}

\newtheorem{prop}{Proposition}
\theoremstyle{definition}

\theoremstyle{remark}
\newtheorem{rmrk}{Remark}
\newtheorem{asmp}{Assumption}

\providecommand{\PLQT}{\textsc{p-lqt}}
\providecommand{\lindata}{\mathcal{L}}       
\providecommand{\quaddata}{\mathcal{Q}}      
\providecommand{\PRIC}{\textsc{p-ric}}       

\IEEEoverridecommandlockouts
\title{\LARGE \bf \socx: GPU-Accelerated Chance-Constrained Stochastic Optimal Control using JAX}

\author{Tanmay Dokania$^{1}$ and Yashwanth Kumar Nakka$^{1}$ 
\thanks{$^{1}$Authors are with the Daniel Guggenheim School of Aerospace Engineering at the Georgia Institute of Technology, Atlanta, GA, 30332 USA. Email: [tdokania3,ynakka3]@gatech.edu
}}
\begin{document}
\maketitle
\thispagestyle{empty}
\pagestyle{empty}
\begin{abstract}
In this article, we present \socx, a GPU-accelerated solver for lifted convex reformulations of linear stochastic optimal control under chance constraints, implemented in JAX. We demonstrate that by exploiting the sparse and independent structure of the problem, distributed optimization can reduce it to simpler problems: independent linear-quadratic tracking (LQT) problems and Euclidean projections onto sets. These simpler problems can be solved in parallel and then coupled using the constraints. To validate our solver, we run it on a hover-linearized 3D quadrotor and show a speed-up of up to $98\times$ over MOSEK and up to $2224\times$ over SCS. \socx\ also solves horizons up to $N=1120$, whereas MOSEK runs out of memory at $N=320$.
We present ablations to isolate the key components that enable the speed-up.
\end{abstract}
\section{Introduction}
Motion planning is a fundamental problem in robotics, as it requires finding an optimal path from the initial to the final state that minimizes length or control effort, is feasible, and avoids obstacles~\cite{lavalle_planning_2006}. Autonomous systems such as unmanned aerial vehicles and spacecrafts, face uncertainties, e.g., wrench perturbations, estimation noise, during tasks like safe exploration~\cite{nakka_chance-constrained_2021,nakka_trajectory_2019}. Therefore, rather than a single path, the problem becomes one of finding a distribution over paths such that the probability of failure is bounded~\cite{nakka_trajectory_2023,yu_gaussian_2023}. 

Stochastic motion planning is an example of stochastic optimal control, and covariance steering solves such problems by designing feedback to control the state distribution, specifically the mean and covariance, from an initial to a final distribution in a finite time horizon. Covariance steering has many applications in robotics and aerospace. However, to the best of the authors' knowledge, analytically optimal results are only known for unconstrained cases~\cite{liu_optimal_2025}. For the constrained cases, numerical optimization methods exist that use feedback parameterization, yielding a nonconvex optimization problem with $O(N n_x (n_u+n_x))$ variables. Computationally efficient formulations approximate the constraints, thereby shrinking the feasible set and yielding suboptimal solutions~\cite{rapakoulias_discrete-time_2023,pilipovsky_computationally_2024}.
Recent developments in lifted representations yield problems that can exactly represent linear chance~\cite{Balci2024ConstrainedMV} and terminal covariance constraints, with convex approximations for quadratic chance constraints~\cite{dokania-lcss2026}, but are computationally prohibitive as the number of variables grows as $O(N^2 n_w (n_u+n_x))$. 

Parallelization introduces a new dimension to algorithm design, enabling new problems to be solved in real time. Batching refers to solving multiple instances of the same problem \cite{turbompc,jaxipm,cusadi}; it increases the number of solves in a given time, but it neither speeds up a single solve nor directly enables solving longer horizons. In contrast, temporal parallelization speeds up a single solve by representing the backward Riccati recursion and the forward rollout as associative scans, reducing their span from $O(N)$ to $O(\log N)$~\cite{sarkka2023}, an approach used in GPU-accelerated iterative linear-quadratic regulator (iLQR) for legged robots~\cite{primal-dual-iLQR}.

\input{gen/lit_table}

The alternating direction method of multipliers (ADMM) is a distributed optimization algorithm that has become the backbone of several GPU-based convex solvers, but they are generic and do not exploit the full optimal control structure. Being relatively immature, they support fewer constraint types. For example, cuOSQP handles only quadratic programs~\cite {cuosqp}, and cuClarabel restricts the dimension of semidefinite blocks~\cite{CuClarabel}. The splitting conic solver (SCS) with GPU linear solver (cuDSS in backend) supports both second-order cone constraints and semidefinite constraints, but solves the generic Karush-Kuhn-Tucker (KKT) system and does not parallelize independent constraint projections~\cite{scs-gpu}.

ADMM has also been specialized for control under uncertainty on GPUs, for robust nonlinear model predictive control~\cite{gpusls} and robust trajectory optimization~\cite{cunrto}. As robust approaches assume bounded disturbances and account for worst-case uncertainty, they lead to overconservatism and infeasibility when the uncertainty is unbounded~\cite{SMPC-mesbah}. This literature gap is further explained in~\Cref{tab:lit_comparison}.

Addressing these limitations, we present \socx: GPU-accelerated algorithm implemented in JAX~\cite{jax2018github} that solves chance-constrained stochastic optimal control by exploiting the lifted convex formulation's structure and achieves minimal conservatism~\cite{dokania-lcss2026}. The key contributions are: (i) a variable split that breaks the problem into independent LQT problems and closed-form parallel projection operations, (ii) an operator split of linear-quadratic tracking (LQT), $\PRIC$ and $\PLQT$ that enables sharing gains across stochastic channels, uses associative scans for efficient parallelization and is 2.5--4.0$\times$ faster per iteration than a sequential LQT, 2.2--7.5$\times$ faster than a condensed KKT (\Cref{tab:sims3d_ablation}), (iii) a small projection operation for the terminal linear matrix inequality (LMI) constraint via an $n_x\times n_x$ Gram eigendecomposition instead of an SVD of the $n_x\times\bas N$ factor or a PSD projection of size $n_x+\bas N$, which is 22--216$\times$ faster per call than the Schur-complement PSD projection, and (iv) validation on 3D linearized quadrotor dynamics demonstrating speed-ups of 7--98$\times$ over MOSEK at $10^{-5}$ ($N=40$--$240$), 47--2224$\times$ faster than SCS+cuDSS. \socx\ solves long-horizon up to $N=1120$, whereas MOSEK runs out of memory at $N=320$ and SCS+cuDSS times out at $N=640$.

We first review the background on the lifted convex formulation of stochastic optimal control in Sec.~II and formulate parallel operators in Sec.~III. Then, \socx\ is presented in Sec.~IV, with simulations and ablations in Sec.~V  followed by the conclusion.

\textit{\textbf{Notation:}} Boldface letters denote random variables (e.g., $\state k, \cntrl k, \basv, \noise k$). $\pd{n}$ ($\psd{n}$) is the cone of $n \times n$ positive definite (semidefinite) matrices, $I_n$ is the identity, and $\quadform{A}{x} = x^\top A x$ for $A \in \pd{n}$. $\Tr{\cdot}$, $\Ex{\cdot}$, and $\Vr{\cdot}$ denote the trace, expectation, and variance. $\N{\mu}{\Sigma}$ is a Gaussian with mean $\mu$ and covariance $\Sigma$, and $\Phi_{\mathcal N}(\cdot)$ and $\Phi_{\chi^2_n}(\cdot)$ are the CDFs of the standard Gaussian and the chi-squared distribution with $n$ degrees of freedom.

\section{Background: Stochastic Control}
We consider the discrete-time stochastic linear time-varying system
\begin{equation}
\begin{split}
\state {k+1} &= A_k \state k + B_k \cntrl k + D_k\noise k, \label{eq:sys_dyn}\\
\state 0 &\sim \N{\mu_0^\star}{\Sigma_0},
\qquad
\noise k \sim \N{0}{I_{n_w}},
\end{split}
\end{equation}
where $\state k\in\R^{n_x}$, $\cntrl k\in\R^{n_u}$, and $\noise k\in\R^{n_w}$ over a horizon $N$.

\begin{asmp}
The initial state and process noises are mutually independent Gaussian random variables,
$\Ex{\noise i\noise j^\top}=\delta_{ij}I_{n_w}$ and
$\Ex{\noise i\state 0^\top}=0$.
The control $\cntrl k$ is causal and affine in
$\state 0,\noise 0,\ldots,\noise{k-1}$; this policy class is denoted
$\Pi_{\mathrm{affine}}$.
\label{asmp:independence_represent}
\end{asmp}

Linear chance constraints on the state and control are
\begin{equation}
\Pr{a_{i,k}^\top\state k\le b_{i,k}}\ge1-\epsilon_{x,i,k},
\qquad
\Pr{\alpha_{i,k}^\top\cntrl k\le\beta_{i,k}}\ge1-\epsilon_{u,i,k},
\label{eq:def_lcc}
\end{equation}
with $a_{i,k}\in\R^{n_x}$, $\alpha_{i,k}\in\R^{n_u}$, $b_{i,k},\beta_{i,k}\in\R$, and $\epsilon_{x,i,k},\epsilon_{u,i,k}\in[0,0.5)$. We additionally consider the terminal quadratic chance constraint
\begin{equation}
\Pr{\quadform{\termQ^{-1}}{(\state N-\Ex{\state N})}\le1}
\ge1-\epsilon_N,
\label{eq:termQ}
\end{equation}
and the terminal covariance bound
\begin{equation}
\Vr{\state N}\preceq\Sigma_f,
\label{eq:cov_control}
\end{equation}
where $\termQ,\Sigma_f\in\pd{n_x}$ and $\epsilon_N\in[0,1]$.

The finite-horizon stochastic optimal control problem is
\begin{problem}[\textbf{Linear Stochastic Optimal Control}]
\label{prob:LSOC}
\begin{equation}
\scalebox{1}{$
\begin{aligned}
\min_{\pi\in\Pi_{\mathrm{affine}}}
J(\state{},\cntrl{})
\quad =\qquad
\sum_{l\in T_{wp}}
\quadform{Q_\star}{(\Ex{\state l}-\mux l^\star)} 
\\
+
\Ex{\sum_{k=0}^{N-1}
\left(
\quadform{Q_k}{\state k}
+
\quadform{R_k}{\cntrl k}
\right)
+ \quadform{Q_N}{(\state N-x^\star)}
} ,
\end{aligned}$}
\label{eq:prob1:orig_cost}
\end{equation}
subject to \eqref{eq:sys_dyn}, \eqref{eq:def_lcc}, \eqref{eq:termQ}, and \eqref{eq:cov_control} for all $k\in\itt{N-1}$.
Here $Q_k,Q_N,Q_\star\succeq0$, $R_k\succ0$, $x^\star\in\R^{n_x}$ is the terminal target, and $\mux l^\star\in\R^{n_x}$ are waypoint targets for $l\in T_{wp}\subseteq\itt{N}$.
\end{problem}

\subsection{Lifted Representation}

Under Assumption~\ref{asmp:independence_represent}, the state and control remain jointly Gaussian. Introduce the independent Gaussian basis
\begin{equation}
\xiv{k+1}:=
\begin{bmatrix}
\xiv k\\
\noise k
\end{bmatrix},
\qquad
\xiv0\sim\N{0}{I_{n_x}},
\label{eq:stoch_basis_def}
\end{equation}
where
$\state0=\mu_0^\star+\mvarx0^\star\xiv0$.
Then $\xiv k\sim\N{0}{I_{\bas k}}$ with
$\bas k:=n_x+kn_w$, and
\begin{subequations}\label{eq:reps}
\begin{align}
\state k &= \mux k+\mvarx k\xiv k,
\label{eq:state_rep}\\
\cntrl k &= \muu k+\mvaru k\xiv k,
\label{eq:control_rep}
\end{align}
\end{subequations}
with $\mux k\in\R^{n_x}$, $\muu k\in\R^{n_u}$, $\mvarx k\in\R^{n_x\times\bas k}$, and $\mvaru k\in\R^{n_u\times\bas k}$. Thus
\begin{equation}
\Ex{\state k}=\mux k,\qquad
\Vr{\state k}=\mvarx k\mvarx k^\top,
\label{eq:state_mean_cov}
\end{equation}
with analogous expressions for $\cntrl k$.

The lifted dynamics are \cite[Lemma~2]{dokania-lcss2026}
\begin{subequations}\label{eq:lifted_dynamics}
\begin{align}
\mux{k+1}
&=A_k\mux k+B_k\muu k,
\label{eq:l2:mu_dynamics}\\
\mvarx{k+1}
&=
\begin{bmatrix}
A_k\mvarx k+B_k\mvaru k & D_k
\end{bmatrix},
\label{eq:l2:ran_dynamics}
\end{align}
\end{subequations}
with $\mux0=\mux0^\star$ and
$\mvarx0=\mvarx0^\star$.

For Gaussian variables, the linear chance constraints admit the exact second-order cone representation
\begin{subequations}\label{eq:lcc_lifted}
\begin{align}
a_{i,k}^\top\mux k
+\zeta_{x,i,k}\|\mvarx k^\top a_{i,k}\|_2
&\le b_{i,k},\\
\alpha_{i,k}^\top\muu k
+\zeta_{u,i,k}\|\mvaru k^\top\alpha_{i,k}\|_2
&\le\beta_{i,k},
\end{align}
\end{subequations}
where
$\zeta_{x,i,k}=\Phi_{\mathcal N}^{-1}(1-\epsilon_{x,i,k})$
and
$\zeta_{u,i,k}=\Phi_{\mathcal N}^{-1}(1-\epsilon_{u,i,k})$.

The terminal covariance bound is equivalent to
\begin{equation}
\mvarx N\mvarx N^\top\preceq\Sigma_f
\iff
\begin{bmatrix}
\Sigma_f & \mvarx N\\
\mvarx N^\top & I_{\bas N}
\end{bmatrix}
\succeq0.
\label{eq:cov_control_lifted}
\end{equation}

The quadratic chance constraint admits the conservative LMI approximation
\begin{equation}
\begin{bmatrix}
\tfrac{1}{\zeta_Q}\termQ & \mvarx N\\
\mvarx N^\top & I_{\bas N}
\end{bmatrix}
\succeq0,
\qquad
\zeta_Q=\Phi_{\chi^2_{n_x}}^{-1}(1-\epsilon_N),
\label{eq:qcc_sdp_lifted}
\end{equation}
or the trace approximation
\begin{equation}
\Tr{\mvarx N^\top\termQ^{-1}\mvarx N}
\le
\frac{1}{
\left(
\Phi_{\mathcal N}^{-1}
\left(
\frac{1+\sqrt[n_x]{1-\epsilon_N}}{2}
\right)
\right)^2}.
\label{eq:qcc_trace_lifted}
\end{equation}
Their relative conservatism is discussed in \cite{dokania-lcss2026}.

\subsection{Convex Reformulation}

Substituting the lifted representation gives the deterministic convex problem
\begin{problem}[\textbf{Lifted Convex Reformulation}]
\label{prob:LDLOC}
\begin{subequations}
\begin{equation}
\scalebox{0.95}{
$\begin{split}
\min_{\mux{},\muu{},\mvarx{},\mvaru{}} J
={}&
\sum_{k=0}^{N-1}
\left(
\quadform{Q_k}{\mux k} + \Tr{Q_k\mvarx k\mvarx k^\top} \right)\\
&+ \sum_{k=0}^{N-1}
\left(
\quadform{R_k}{\muu k}
+
\Tr{R_k\mvaru k\mvaru k^\top}
\right)
\\
&+
\quadform{Q_N}{(\mux N-x^\star)}
+
\Tr{Q_N\mvarx N\mvarx N^\top}\\ &
+
\sum_{i\in T_{wp}}
\quadform{Q_\star}{(\mux i-\mux i^\star)} .
\end{split}$}
\label{eq:thm1:cost}
\end{equation}
\begin{align}
\text{s.t.}\quad
&\eqref{eq:lifted_dynamics},\ \forall\ k\in\itt{N-1},
\label{eq:thm1:sys-dyn}\\
&\eqref{eq:lcc_lifted},\ \forall\ i,\ \forall\ k\in\itt{N},
\label{eq:thm1:lcc-lifted}\\
&\eqref{eq:cov_control_lifted},
\quad
\eqref{eq:qcc_sdp_lifted}
\ \text{or}\
\eqref{eq:qcc_trace_lifted}.
\label{eq:thm1:qcc_cov}
\end{align}
\label{eq:prob:lifted}
\end{subequations}
\end{problem}

\begin{theorem}[\textbf{Exact Convex Reformulation}]
\cite[Theorem~1]{dokania-lcss2026}
\label{thm:exact_convex_reformulation}
Under Assumption~\ref{asmp:independence_represent},
\Cref{prob:LDLOC}, restricted to
\eqref{eq:lifted_dynamics},
\eqref{eq:lcc_lifted}, and
\eqref{eq:cov_control_lifted},
is an exact deterministic convex reformulation of
\Cref{prob:LSOC}: the two problems have equivalent costs and feasible affine policies.
\end{theorem}

\begin{prop}[\textbf{Conservative Approximation of QCCs}]
\cite[Proposition~1]{dokania-lcss2026}
\label{prop:conservative_qcc}
Both \eqref{eq:qcc_sdp_lifted} and
\eqref{eq:qcc_trace_lifted} are sufficient approximations of
\eqref{eq:termQ}; neither uniformly dominates the other.
\end{prop}

\section{Background: Parallel Primitives}
In this section, we review parallel primitives that are packaged as an operator, $\PLQT$, for solving the Linear-Quadratic Tracking (LQT) problem using an associative-scan that evaluates in $O(\log N)$ span that is invoked for mean and stochastic channels in \Cref{sec:method}.

\subsection{Parallel Associative Scan}\label{sec:assoc-scan}
The associative property states that the grouping of operations does not matter, i.e., for an operator $\mathbin{{\bullet}}$, $a\mathbin{{\bullet}}(b\mathbin{{\bullet}} c) = (a\mathbin{{\bullet}} b) \mathbin{{\bullet}} c$ holds. Defining associative operations allows to compute $(a_1,\ a_1\mathbin{{\bullet}} a_2,\ \dots,\ a_1\mathbin{{\bullet}}\cdots\mathbin{{\bullet}} a_n)$ from $(a_1,\dots,a_N)$ in $O(\log N)$ span as opposed to a sequential recursion in $O(N)$ span~\cite{blelloch1989}. 

To compute the rollouts of a dynamical system with specified feedback matrices, the following recursion is defined: $y_{k+1} = M_k y_k + v_k$, called an affine recurrence. This can be expressed as an associative operator, with the $k$-th element being $(M_k, v_k)$, and the operation being

\begin{equation}
    (M_b, v_b) \bullet (M_a, v_a) = (M_b M_a,M_b v_a+v_b). \label{eq:affine-recurrance}
\end{equation}
For prediction horizon $N$, this enables the computation of {rollouts} in $O(\log N)$ {span}, instead of $O(N)$. The price is paid because it requires additional operations and memory to store the intermediate matrix~\cite{sarkka2023,primal-dual-iLQR}. 

\subsection{The Parallel LQT Operator}\label{sec:plqt}
A finite-horizon LQT problem is defined as
\begin{equation}
\begin{aligned}
    \min_{\chu}\ &\sum_{k=k_0}^{N}\big(\quadform{\augQ_k}{\chx_k}-2\augq_k^\top\chx_k\big)
    + \sum_{k=k_0}^{N-1}\big(\quadform{\augR_k}{\chu_k}-2\augr_k^\top\chu_k\big) \\
    \text{s.t.}\ &\ \chx_{k+1}=A_k\chx_k+B_k\chu_k, \qquad \chx_{k_0} \text{ given},
\end{aligned}
\label{eq:channel_lqr}
\end{equation}
with $N, k_0, \chx, \chu, A, B$ being the horizon, start step, state, control input, and transition matrices. And $\augQ \succeq 0, \augR \succ 0$ denote the quadratic costs for state and control with linear terms $\augq, \augr$.
Therefore, a problem is fully specified by quadratic $\quaddata$  and linear parameters $\lindata$,

\begin{equation}
\begin{aligned}
    \quaddata &:= \big(A_{0:N-1},\,B_{0:N-1},\,\augQ_{0:N},\,\augR_{0:N-1}\big),\\
    \lindata &:= \big(k_0,\,\chx_{k_0},\,\augq_{k_0:N},\,\augr_{k_0:N-1}\big),
\end{aligned}
\label{eq:lqt_data}
\end{equation}
here, the subscript $i:j$ denotes the set of matrices with the index varying from $i$ to $j$.
The optimal law can be represented as an affine-feedback law, $\chu_k = K_k\chx_k+\kff_k$.
Demonstrated in \cite{sarkka2023}, $\chu_k$ can be computed from $(\quaddata,
\lindata)$ in $O(\log N)$ using an parallel associative scan. We will summarize only the properties and steps that we will use later.

We denote the value function at time step $k$, the minimum cost of starting at state $\chx$ at time $k$, as
\begin{equation}
    \Vfun_k(\chx) = \quadform{P_k}{\chx} - 2p_k^\top\chx + \text{const}.
    \label{eq:value_fn}
\end{equation}

Firstly, $P_k$ can be obtained from $\quaddata$ using associative scan~\cite{primal-dual-iLQR} and then the feedback matrix, $K_k$ can be constructed using the relation:
\begin{equation}
    K_k = -G_k^{-1}B_k^\top P_{k+1}A_k, \quad G_k := \augR_k + B_k^\top P_{k+1}B_k.
    \label{eq:kfb}
\end{equation}
We package this as an operator, $\PRIC$, which has $O(\log N)$ span, $O\big(N(n_x^3+n_u^3)\big)$ work and $O(Nn_x^2)$ memory,
\begin{equation}
    \big(P_{0:N},\,K_{0:N-1}\big) = \PRIC\big(\quaddata\big).
    \label{eq:pric_operator}
\end{equation}
Additionally, following Bellman optimality, $(P_k,K_k)$ depend only on $(A,B,\augQ,\augR)_{k:N}$, so an instance that starts at $k_0>0$ can use the same solve.

Using the feedback matrix, the linear term of the value function, $p_k$, can be computed using $\lindata$ as the recursion
\begin{equation}
    p_k=(A_k+B_kK_k)^\top p_{k+1}+\augq_k+K_k^\top\augr_k,
    \label{eq:pk}
\end{equation}
which is an affine recurrence relation in $p_{k}$, with $p_N=\augq_N$. This can be obtained in $O(\log N)$ as defined in \Cref{sec:assoc-scan}.
The feedforward input $\kff$ can be computed as
\begin{equation}
    \kff_k = G_k^{-1}\big(\augr_k + B_k^\top p_{k+1}\big),
    \label{eq:kff}
\end{equation}
Now, as the control input is $\chu_k = K_k \chx_k + \kff_k$ and the optimal states can be defined using the affine recurrence relation, as the initial state is given:
\begin{equation}
    \chx_{k+1} = (A_k + B_k K_k) \chx_k + B_k \kff_k .
    \label{eq:rollout}
\end{equation}
Combining the two affine recurrences \eqref{eq:pk} and \eqref{eq:rollout}, we package it as an operator:
\begin{equation}
    \big(\chx_{k_0:N},\,\chu_{k_0:N-1}\big) = \PLQT\big(\quaddata,\,P,\,K,\,\lindata\big ).
    \label{eq:plqt_ff_operator}
\end{equation}
For fixed $(\quaddata,P,K)$, the map $\lindata\mapsto(\chx,\chu)$ given by \eqref{eq:pk}, \eqref{eq:kff}, \eqref{eq:rollout} is \emph{linear} in $(\chx_{k_0},\augq,\augr)$.
If it is invoked $W$ times, in parallel, with shared $(\quaddata,P,K)$ then it has $O(\log N)$ span, $O\big(Nn_x^3+WNn_x^2\big)$ work and $O(WNn_x)$ memory.

\input{algo_diag}

\section{Method}
\label{sec:method}
In this section, we present {\socx} by describing (i) the structure of \Cref{prob:LDLOC}; (ii) ADMM with the variable split; (iii) the $\admmx$-update using $\PRIC$/$\PLQT$ operations; and (iv) the $z$-update as closed-form projections for second-order cone and covariance constraint, as summarized in \Cref{fig:pipeline}.
The resulting solver, \socx, does not require Jacobians, KKT matrices, or dense constraint operators.

\subsection{Structure of the Lifted Problem}
The number of variables in \Cref{prob:LDLOC} grows as $O(N^2)$ because of the new elements in the stochastic basis. However, the problem is highly structured.

\emph{Shared dynamics.} Let $\colx^{(c)}_k := \mvarx k[:,c]$ and $\colu^{(c)}_k := \mvaru k[:,c]$ denote the columns of $\mvarx{}$ and $\mvaru{}$. The following equations govern the stochastic columns $(\colx^{(c)}, \colu^{(c)})$, obtained from~\eqref{eq:l2:ran_dynamics}:
\begin{equation}
\begin{aligned}
    &\colx^{(c)}_{k+1} = A_k \colx^{(c)}_k + B_k \colu^{(c)}_k,& k\ge\birth c,\\
    &\colx^{(c)}_k = 0,\quad \colu^{(c)}_k = 0,& k<\birth c.
\end{aligned}
    \label{eq:column_decoupling}
\end{equation}
The initial conditions are defined as:
\begin{equation}
\scalebox{0.97}{$
    \big(\birth c,\colx^{(c)}_{\birth c}\big) =
    \begin{cases}
        \big(0,\ \mvarx 0^\star[:,c]\big), & c<n_x,\\
        \big(\lfloor \frac{(c-n_x)}{n_w}\rfloor+1, D_{\birth c-1}[:,r_c]\big), & c\ge n_x,
    \end{cases}
    \label{eq:birth}
$}
\end{equation}
with $r_c:=(c-n_x)\bmod n_w$. Similarly, the mean dynamics are independent of stochastic columns~\eqref{eq:l2:mu_dynamics}.
    
\emph{Cost separation.} In the lifted cost function \eqref{eq:thm1:cost}, the stochastic columns appear through the same matrices, $Q_k$ and $R_k$, and there are no coupling terms across columns. $\Tr{Q_k\mvarx k\mvarx k^\top} = \sum_c
    \quadform{Q_k}{\colx^{(c)}_k}$, and likewise for $R_k$ and $\mvaru k$. The mean variables aren't coupled with the stochastic columns in the lifted cost function.

\emph{Coupling constraints.} Second-order cone constraints couple the mean and stochastic channels~\eqref{eq:thm1:lcc-lifted}. The quadratic and the linear matrix inequality constraint couples the stochastic columns~\eqref{eq:thm1:qcc_cov}.

Dynamics and cost are separable and coupled only through the constraints; therefore, this motivates an optimization algorithm that allows solving subproblems and pooling using the constraints. This structure is what ADMM is designed to exploit when an appropriate variable split is used.

\subsection{ADMM \& Variable Splitting}\label{sec:admm}
We review ADMM, a popular algorithm for distributed convex optimization, and then define a variable split. ADMM is well-suited to large-scale problems and combines decomposability with superior convergence properties \cite{boyd-admm-2011}. It solves problems of the form
\begin{subequations}\label{eq:admm_generic_obj}
\begin{align}
    \min_{\admmx,z} \quad & f(\admmx) + g(z) \\ 
    \text{s.t.} \quad& \Mop  \admmx+ {\mathcal E} z = c 
\end{align}
\end{subequations}
by formulating the augmented Lagrangian $L_\rho (\admmx,z,y) = f(\admmx) + g(z) + y^\top (\Mop \admmx+{\mathcal E} z-c) + \frac\rho2 \|\Mop \admmx+{\mathcal E} z-c \|_2^2$ {, and iterating over the update steps}
\begin{subequations}
\begin{align}
    \admmx\text{-update}&: \admmx^{{j}+1} := \argmin_{\admmx} L_\rho (\admmx, z^{ j}, y^{ j}) \label{eq:admm-x-update}\\
    z\text{-update} &: z^{ j+1} := \argmin_{z} L_\rho (\admmx^{ j+1}, z, y^{ j}) \label{eq:admm-z-update}\\
    y\text{-update}& : y^{ j+1} := y^{ j} + \rho (\Mop \admmx^{ j+1} + {\mathcal E} z^{ j+1}-c) \label{eq:admm-y-update}
\end{align}\label{eq:admm_def}
\end{subequations}
with $\rho>0$.

Introducing the scaled dual variable $\admmu:= y/\rho$ and completing the squares, the last two terms of the augmented Lagrangian, for the case ${\mathcal E} = -I$, become
\begin{equation}
L_\rho (\admmx, z,\rho \admmu) = f(\admmx) + g(z) + \frac \rho 2 \|\Mop \admmx - z +\admmu \|_2^2 - \frac \rho 2 \|\admmu \|_2^2 \label{eq:scaled_lagrangian}
\end{equation}
This squared penalty term turns the $\admmx$-update into an LQT \eqref{eq:channel_lqr}.

{\emph{Variable splitting.}} Collect the primal variables as $$\admmx := \left(\mux{0:N}, \muu{0:N-1}, \mvarx{0:N}, \mvaru{0:N-1}\right).$$
{For each cone constraint, we introduce a split variable whose size matches the cone.} Let $z_{x,i,k},z_{u,i,k}  \in \R^{1+\bas{k}}$ correspond to the state and control cone constraint, defined as
\begin{equation}
\begin{split}
z_{x,i,k} = \Mop_{x,i,k} \admmx &:= \big(b_{i,k}-a_{i,k}^\top\mux k,\ \zeta_{x,i,k}\mvarx k^\top a_{i,k}\big) \\
z_{u,i,k} =  \Mop_{u,i,k} \admmx &:= \big(\beta_{i,k}-\alpha_{i,k}^\top\muu k,\ \zeta_{u,i,k}\mvaru k^\top \alpha_{i,k}\big).
\end{split}\label{eq:admm-lcc}
\end{equation}

Additionally, depending on the terminal constraint, either \eqref{eq:cov_control_lifted} or \eqref{eq:qcc_sdp_lifted}, the constraint takes the form $\mvarx N\mvarx N^\top \preceq \covub$, with $\covub = \Sigma_f$ or $\covub = \zeta_Q^{-1}\termQ$ respectively. 
\begin{rmrk}
    The discussion is limited to the linear matrix inequality as the trace relaxation \eqref{eq:qcc_trace_lifted} appears as a quadratic constraint that can be expressed as second-order cone constraint.
\end{rmrk}
Hence, the corresponding split variable $z_N\in \R^{n_x \times \bas{N}}$ is defined as
$z_N = \Mop_N \admmx := \covub^{-1/2}\mvarx N$.
Collecting the split variables, $z:=\{z_{x,i,k}, z_{u,i,k}, z_N\}$, an affine operator can be defined as $\Mop \admmx = z$. 

Splitting methods handle a convex constraint set $\mathcal C$ by moving it out of the smooth objective and into the indicator function
\begin{equation}
    \ind{\mathcal C}(z) = \begin{cases} 0, & z \in \mathcal C\\ +\infty, & z \notin \mathcal C \end{cases},
    \label{eq:indicator}
\end{equation}
which is convex and closed whenever $\mathcal C$ is.
Hence, resulting in the following ADMM split of \Cref{prob:LDLOC}:
\begin{equation}
\begin{aligned}
    f(\admmx) &= J(\admmx) + \ind{\text{\eqref{eq:lifted_dynamics}}}(\admmx), \\
    g(z) &= \sum_{i,k} \ind{\Ksoc_{\bas k}}(z_{x,i,k}) + \sum_{i,k} \ind{\Ksoc_{\bas k}}(z_{u,i,k}) + \ind{\Bspec}(z_N), \\
    \text{s.t. }&\Mop \admmx - z = 0
    \label{eq:fg_split}
\end{aligned}
\end{equation}
where $\Ksoc_n = {\{(t,s)\in\R\times\R^{n} : \|s\|_2\le t\}}$ for the second-order cone constraints and  $\Bspec := \{Z\in \R^{n_x \times \bas{N}} : \|Z\|_2\le 1\}$ for the terminal covariance constraint. Under this split, the $\admmx$-update becomes independent LQT solves and the $z$-update becomes closed-form set projections as described in the next subsections.

\emph{Convergence monitoring.} The primal $(r_p^j)$ and dual residuals $(r_d^j)$ are defined as
\begin{equation}
    r_p^{j} = \|\Mop \admmx^{j} + {\mathcal E} z^{j} - c\|_2, \
    r_d^{j} = \rho\|\Mop^\top {\mathcal E}(z^{j} - z^{j-1})\|_2 .
    \label{eq:admm_residuals}
\end{equation}
With an initialization of $z^0=0$ and $\admmu^0=0$, \socx\ monitors the following two conditions and stops when both hold
\begin{equation}
\begin{aligned}
    \|\Mop\admmx^{j}-z^{j}\|_\infty &\le \epsilon_{\mathrm{abs}}+\epsilon_{\mathrm{rel}}\max\big(\|\Mop\admmx^{j}\|_\infty,\|z^{j}\|_\infty\big),\\
    \rho\|z^{j}-z^{j-1}\|_\infty &\le \epsilon_{\mathrm{abs}}+\epsilon_{\mathrm{rel}}\,\rho\|\admmu^{j}\|_\infty,
\end{aligned}
\label{eq:stop}
\end{equation}
where $\epsilon_{\mathrm{rel}}$ and $\epsilon_{\mathrm{abs}}$ are the relative and absolute tolerance parameters \cite{cuosqp}.

The value of $\rho$ determines the convergence rate, and adaptive algorithms help reduce the number of iterations by scaling $\rho$ based on the residuals \cite{gpusls,boyd-admm-2011}
\begin{equation} \label{eq:rho_balance}
\rho^+ \leftarrow \rho \cdot \operatorname{clip}\left(\sqrt{{r_p^{ j}}/{{ r_d^{\,j}}}},\ \underline\kappa,\ \overline\kappa\right),\quad  \admmu^+ \leftarrow  \frac{\admmu\rho}{\rho^+}.
\end{equation}
thereby eliminating the need to hand-tune $\rho$ and driving the residuals to a comparable magnitude, with the safeguards $\underline\kappa = 1/2$, $\overline\kappa = 2$.

\begin{rmrk}[{Over-relaxation}]\label{rmrk:overrelaxation}
    {A standard empirical accelerator is} over-relaxation, which substitutes $\Mop \admmx^{j+1}$ with ${\alpha_{\mathrm{or}}} \Mop \admmx^{j+1} + (1-{\alpha_{\mathrm{or}}})(c-{\mathcal E} z^j)$ in the $z$- and $\admmu$-updates. Convergence is retained for ${\alpha_{\mathrm{or}}}\in(0,2)$, and ${\alpha_{\mathrm{or}}}\in[1.5,1.8]$ improves the linear rate at no per-iteration cost \cite{boyd-admm-2011}. 
\end{rmrk}

\subsection{$\admmx$-update as Independent LQT Invocations}\label{sec:x-update}

Under the split \eqref{eq:fg_split}, the $\admmx$-update step \eqref{eq:admm-x-update} keeps the lifted dynamics \eqref{eq:lifted_dynamics} as equality constraint,
\begin{equation}
\begin{split}
    \admmx^{j+1} = \argmin_{\admmx} \ &J(\admmx) + \frac{\rho}{2} \|\Mop \admmx - \tgt \|_2^2 \\
    \text{s.t.} \ & \eqref{eq:lifted_dynamics}
\end{split} \label{eq:primal-update}
\end{equation}
here, $\tgt := z^j-\admmu^j$. 
For all the columns, the added penalty term is a sum of squares of components of $\Mop \admmx$ that acts on $(\mux{}, \muu{}, \mvarx{}, \mvaru{})$ column-wise, without any cross-column coupling. Hence, $\admmx$-update \eqref{eq:primal-update} splits into $1+\bas N$ independent instances of LQT solves~\eqref{eq:channel_lqr}, consisting of one {mean channel} $(\chx,\chu)\mapsto(\mux{},\muu{})$ and $\bas N$ {stochastic channels} $(\chx,\chu)\mapsto(\colx^{(c)},\colu^{(c)})$.

Hence, the defined operators can be used to solve this step with the arguments $(\quaddata,\lindata)$ described in this subsection and summarized in \Cref{tab:channels}.

{\emph{Quadratic data.}} Collecting the base cost and the $\rho$-penalty gives the augmented stage weights
\begin{subequations}\label{eq:augmented_weights}
\begin{align}
    \augQ^{\fmean}_k &:= Q_k + \tfrac\rho2\textstyle\sum_i a_{i,k}a_{i,k}^\top
      + \boldsymbol{1}_{[k\in T_{wp}]}\,Q_\star, \\
    \augR^{\fmean}_k &:= R_k + \tfrac\rho2\textstyle\sum_i \alpha_{i,k}\alpha_{i,k}^\top,\\
    \augQ^{\colx}_k &:= Q_k + \tfrac\rho2\textstyle\sum_i \zeta_{x,i,k}^2\, a_{i,k}a_{i,k}^\top, \\
    \augR^{\colx}_k &:= R_k + \tfrac\rho2\textstyle\sum_i \zeta_{u,i,k}^2\, \alpha_{i,k}\alpha_{i,k}^\top, \\
    {\augQ^{\fmean}_N} &{:= Q_N + \tfrac\rho2\textstyle\sum_i a_{i,N}a_{i,N}^\top
      + \boldsymbol{1}_{[N\in T_{wp}]}\,Q_\star,} \\
    \augQ^{\colx}_N &:= Q_N + \tfrac\rho2\textstyle\sum_i \zeta_{x,i,N}^2\, a_{i,N}a_{i,N}^\top
    + \tfrac\rho2 \covub^{-1} .
\end{align}
\end{subequations}
Here, $\boldsymbol{1}$ is an indicator function which is $1$  if $k\in T_{wp}$, otherwise $0$. And $\augQ^{\fmean},\augQ^{\colx} (\augR^{\fmean},\augR^{\colx})$ are the quadratic costs for the state (control) in the mean and stochastic channels. They differ in the quadratic costs because of the different way the chance constraints show up and due to the waypoint costs on the mean. Additionally, the terminal covariance constraint only shows up for the stochastic channels. 

{\emph{Linear data.}} The ADMM target $\tgt$ enters only through linear terms, and differs across each of the stochastic channel. Let the scalar and vector parts of the cone $(i,k)$ be denoted by $(\tgtsc_{i,k},\tgtvc_{i,k})$, and let $\tgtball$ be the target of the terminal ball. To obtain the linear terms, $\augq, \augr$, the penalty in \eqref{eq:primal-update} together with the two reference-tracking terms of \eqref{eq:thm1:cost} is expanded:
\begin{subequations}\label{eq:linear_terms}
\begin{align}
    \augq^{\fmean}_k := &\tfrac\rho2\textstyle\sum_i (b_{i,k}-\tgtsc_{x,i,k})a_{i,k}
    + \delta_{kN}\,Q_N x^\star \nonumber\\
    &+ \boldsymbol{1}_{[k\in T_{wp}]}\,Q_\star\mux k^\star,\\
    \augr^{\fmean}_k := &\tfrac\rho2\textstyle\sum_i (\beta_{i,k}-\tgtsc_{u,i,k})\alpha_{i,k},\\
    \augq^{\colx,(c)}_k := &\tfrac\rho2\textstyle\sum_i \zeta_{x,i,k}\,\tgtvc_{x,i,k,c}\,a_{i,k}
    + \delta_{kN}\,\tfrac\rho2 \covub^{-1/2} \tgtball_{:,c},\\
    \augr^{\colx,(c)}_k := &\tfrac\rho2\textstyle\sum_i \zeta_{u,i,k}\,\tgtvc_{u,i,k,c}\,\alpha_{i,k},
    \qquad k\ge\birth c,
\end{align}
\end{subequations}

{\emph{Dynamics data.}} 
The dynamics equations are shared as described for the stochastic channels in~\eqref{eq:column_decoupling} with different initial conditions~\eqref{eq:birth}, and for the mean channels as defined in \eqref{eq:l2:mu_dynamics}.

\emph{Invocation.} The first step in $\admmx$-update is evaluated as parallel Ricatti scans:
\begin{subequations}
\begin{align}
    (P^{\fmean},K^{\fmean}) &\leftarrow \PRIC\big(\quaddata^{\fmean}\big), \label{eq:pric-mu-invoke}\\
    (P^{\colx},K^{\colx}) &\leftarrow \PRIC\big(\quaddata^{\colx}\big), \label{eq:pric-var-invoke}
\end{align}\label{eq:plqt_calls_ric}
\end{subequations}
followed by affine recurrances scans that use the same feedback gains for the stochastic channels:
\begin{subequations}
\begin{align}
    (\mux{},\muu{}) &\leftarrow \PLQT\big(\quaddata^{\fmean},P^{\fmean},K^{\fmean},\lindata^{\fmean}\big),
    \label{eq:plqt_calls_mean}\\
    (\colx^{(c)},\colu^{(c)}) &\leftarrow \PLQT\big(\quaddata^{\colx},P^{\colx},K^{\colx},\lindata^{(c)}\big),\\
    &c \in \itt{\bas N -1}, \notag
    \label{eq:plqt_calls_cov}
\end{align}\label{eq:plqt_calls_ric}
\end{subequations}

where \eqref{eq:plqt_calls_ric} is evaluated only when $\rho$ changes, and
\begin{equation}
\begin{aligned}
    \quaddata^{\fmean} &:= \big(A,\,B,\,\augQ^{\fmean},\,\augR^{\fmean}\big),\quad
    \quaddata^{\colx} := \big(A,\,B,\,\augQ^{\colx},\,\augR^{\colx}\big),\\
    \lindata^{\fmean} &:= \big(0,\ \mux 0^\star,\ \augq^{\fmean}_{0:N},\ \augr^{\fmean}_{0:N-1}\big),\\
    \lindata^{(c)} &:= \big(\birth c,\ \colx^{(c)}_{\birth c},\ \augq^{\colx,(c)}_{\birth c:N},\ \augr^{\colx,(c)}_{\birth c:N-1}\big).
\end{aligned}
\label{eq:channel_data}
\end{equation}
Hence, \socx\ solves LQT many times with shared gains, rather than formulating a single large conic program.
\begin{table}[t]
\centering
\caption{$\PRIC$ and $\PLQT$ arguments with shared $(A_k,B_k)$.}
\label{tab:channels}
\begin{tabular}{lll}
\toprule
 & mean channel & stochastic channel $c$ \\
\midrule
$(\chx_k,\chu_k)$ & $(\mux k,\muu k)$ & $(\colx^{(c)}_k,\colu^{(c)}_k)$ \\
$\quaddata$: $(\augQ_k,\augR_k)$ & $(\augQ^{\fmean}_k,\augR^{\fmean}_k)$ & $(\augQ^{\colx}_k,\augR^{\colx}_k)$ \\
$\lindata$: $k_0$ & $0$ & $\birth c$ \eqref{eq:birth} \\
$\lindata$: $\chx_{k_0}$ & $\mux 0^\star$ & $\colx^{(c)}_{\birth c}$ \eqref{eq:birth} \\
$\lindata$: $(\augq_k,\augr_k)$ & $(\augq^{\fmean}_k,\augr^{\fmean}_k)$ & $(\augq^{\colx,(c)}_k,\augr^{\colx,(c)}_k)$ \\
\midrule
count & $1$ & $\bas N$ \\
$\PRIC$ calls / $\rho$ update & $1$ & $1$, shared \\
$\PLQT$ calls / iteration & $1$ & $\bas N$, batched \\
\bottomrule
\end{tabular}
\end{table}

\subsection{$z$-update as Parallel Projections}\label{sec:constraint-proj}
In the minimization problem of $z$-update, when the cost $g$ \eqref{eq:fg_split} is a sum of indicators, the update reduces to the proximal operator of the individual indicator, i.e.\ the Euclidean projection,
\begin{equation}
\begin{split}
    z^{j+1} &= \argmin_{z\in\mathcal C} \tfrac12\|z - (\Mop \admmx^{j+1} + \admmu^j)\|_2^2\\
    &=: \proj{\mathcal C}\!\left(\Mop \admmx^{j+1}+\admmu^j\right).
    \label{eq:prox_is_projection}
\end{split}
\end{equation}
 The projections are executed in parallel because the sets are independent; i.e., no variable is projected in both sets simultaneously.

{\emph{Second-order cone constraints.}} For the projection to the set $\Ksoc_{n}$, we obtain
\begin{equation}
    \proj{\Ksoc_n}({t,s}) =
    \begin{cases}
        {(t,s)}, & \|s\|_2 \le t,\\[2pt]
        (0,0), & \|s\|_2 \le -t,\\[2pt]
        \dfrac{t+\|s\|_2}{2}{\left(1,\, \dfrac{s}{\|s\|_2}\right)}, & \text{otherwise.}
    \end{cases}
    \label{eq:soc_projection}
\end{equation}
Therefore, each of $z_{x,i,k}$, similarly for $z_{u,i,k}$, is updated as:
\begin{equation}
    z_{x,i,k}^{j+1} = \proj{\Ksoc_{\bas k}} (\Mop_{x,i,k} \admmx^{j+1} + \admmu_{x,i,k}^{j})
\end{equation}
here, $\Mop_{x,i,k}$ is as defined in \eqref{eq:admm-lcc} and $\admmu_{x,i,k}$ is the corresponding component of $\admmu$.

{\emph{Terminal covariance constraint.}} For projection of $X\in \R^{n_x \times \bas N}$ to the set $\Bspec$, the standard approach is computing the singular value decomposition of ${X = U_X\Lambda V_X^\top}$ and then
${\proj{\Bspec}(X) = U_X \min (\Lambda, I_{n_x}) V_X^\top}$ with $\min$ denoting an element-wise minimum of the two diagonal matrices.
However, as $X$ is a fat matrix and we don't require the matrix $V_X$, we can instead compute the eigendecomposition of ${XX^\top = U_X\Lambda^2U_X^\top}$, giving us ${U_X}$ and ${\Lambda^2}$, from which we construct ${M_X:=U_X\min(I_{n_x}, \Lambda^{-1})U_X^\top}$. Therefore, the projection operator becomes:

\begin{equation}
\begin{aligned}
    \proj{\Bspec}(X) &=  U_X \min (\Lambda, I_{n_x}) V_X^\top \\
    =&  U_X\min(I_{n_x}, \Lambda^{-1})(U_X^\top U_X) \Lambda V_X^\top = M_X X .
\end{aligned}
\label{eq:gram_identity}
\end{equation}

Hence, the $z_N$ update is
\begin{equation}
    z_N^{j+1} = \proj{\Bspec}(\Mop_N \admmx^{j+1}+\admmu^j_N)  \label{eq:gram_projection}
\end{equation}

Hence, the $z$-update reduces to closed-form Euclidean projections onto the cone and covariance constraint considered, which can be evaluated in a parallel, batched process.

\begin{figure*}[t]
  \centering
  \begin{subfigure}[b]{0.362\textwidth}
    \includegraphics[width=\linewidth]{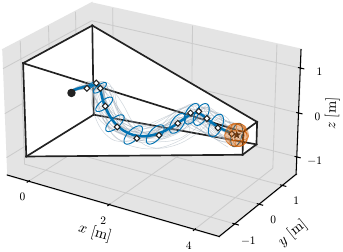}
    \caption{Helix through a tapering corridor}\label{fig:traj3d-helix}
  \end{subfigure}\hfill
  \begin{subfigure}[b]{0.310\textwidth}
    \includegraphics[width=\linewidth]{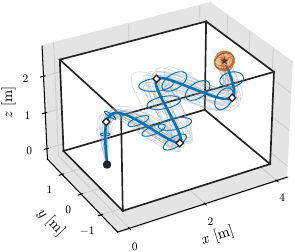}
    \caption{Zig-zag climb through a room}\label{fig:traj3d-zigzag}
  \end{subfigure}\hfill
  \begin{subfigure}[b]{0.308\textwidth}
    \includegraphics[width=\linewidth]{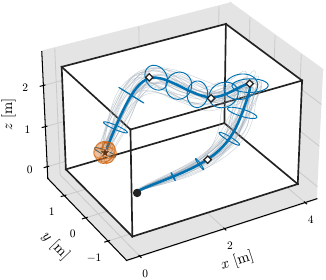}
    \caption{Out and back through a room}\label{fig:traj3d-outback}
  \end{subfigure}\\[2pt]
  \includegraphics[width=0.784\textwidth]{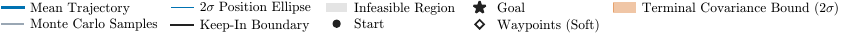}
  \caption{3D quadrotor trajectories computed with \socx, with random wrench perturbations, chance constraints on position \& actuator limits and terminal covariance constraint $(n_x=12, n_w=6, n_u=4, N=80, \Delta t =0.1)$.
  }
  \label{fig:traj3d}
\end{figure*}
\section{Numerical Simulations}\label{sec:sims}
In this section, we present numerical simulations with ablations, run on a system with an Intel Core Ultra 9 285K (64GB RAM) and an RTX 5080 (16GB VRAM).
To demonstrate the efficacy of \socx, we consider a hover-linearized quadrotor with state $(p,v,\phi,\theta,\psi,\omega)\in\R^{12}$, inputs $(\Delta T,\tau_x,\tau_y,\tau_z)\in\R^4$, and noise $\noise k\sim\N{0}{I_6}$ entering as forces and torques, with Euler discretization and $\Delta t=0.1$s. We consider three different setups: a helix in a tapering corridor (13 waypoints), a zig-zag climb, and an out-and-back (4 waypoints each), shown in \Cref{fig:traj3d}. At $N=40$, we reduce the waypoints to 3, 2, and 2, respectively, as the 4-s horizon is too short to fly through all of them.

The walls are represented as six half-plane linear chance constraints \eqref{eq:def_lcc} in position $\epsilon_x=0.05$, and actuator limits are imposed as both an upper and lower limit with $\epsilon_u=0.05$ at each time step. The terminal state has a position-covariance constraint \eqref{eq:cov_control} with $\Sigma_f=(0.12\,\mathrm{m})^2 I_3$. A selector matrix of appropriate size can be introduced to impose such a terminal covariance constraint. A quadratic penalty on state and control input is imposed. The waypoint and goal terms act only on the mean state.

\subsection{{\socx} vs. MOSEK vs. SCS+cuDSS: Tolerance vs. Solve Time}

\begin{figure}[t]
  \centering
  \includegraphics[width=\columnwidth]{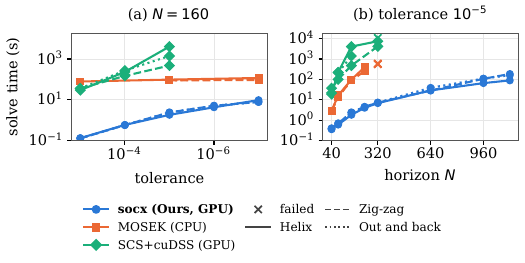}
  \caption{Solve time vs. (a) tolerance at fixed $N=160$ and (b) horizon at fixed $\epsilon=10^{-5}$}
  \label{fig:sims3d}
\end{figure}

For baseline comparisons, we report the internal solve times of MOSEK~\cite{mosek} and SCS (cuDSS), both called through CVXPY~\cite{diamond2016cvxpy}. \socx's measured time excludes JAX's just-in-time (JIT) compilation~\cite{jax2018github}.
\Cref{fig:sims3d} and \Cref{tab:sims3d_main} compare the solve times. Compared to MOSEK, the speed-up grows from 7.2--7.9$\times$ at $N=40$ to 57--98$\times$ at $N=240$ because MOSEK's time grows faster than \socx. MOSEK runs out of memory at $N=320$, while \socx\ solves $N=1120$ in 145\,s on average. As MOSEK is an interior-point method, its solve time does not vary much with tolerance, unlike \socx\ which is a first-order method. We observe that \socx\ remains faster at tight tolerances, but by a smaller margin. At $N=160$ the speed-up is 37--53$\times$ at $10^{-5}$ and still 12--14$\times$ at $10^{-7}$.
SCS (cuDSS), a GPU-based first-order method, is 47--2224$\times$ slower than \socx\ and slower than MOSEK for $N\ge 80$. We additionally observe that SCS times out at a 10,000s time cap for $N\ge640$. In \Cref{tab:sims3d_main}, $\frac{P_\mathrm{viol}}{\epsilon}$ denotes the probability of violation divided by the allowed violation. A value close to 1 indicates that the constraints are active, and a value greater than 1 quantifies the violation.
\input{gen/sims3d_main}
\subsection{$\admmx$-update: LQT ablations}
\Cref{tab:sims3d_ablation} compares the per-iteration cost of the parallel LQT update with three alternatives.
Sequential Riccati updates have $O(N)$ span and use 2.5--4.9$\times$ less memory, but are 2.5--4.0$\times$ slower. The condensed KKT update uses 3.9--9.5$\times$ less memory but is 2.2--7.5$\times$ slower, and its time grows $3.9\times$ and $6.4\times$ per doubling of $N$.
The key comparison is with a per-column loop that does not exploit the fact that the stochastic columns share $\quaddata$ and therefore $(P, K)$, so the gains cannot be reused \eqref{eq:pric-var-invoke}. It is 6.9--14.8$\times$ slower and uses 4.8--9.7$\times$ more memory than the shared-gain update, and both gaps widen as $N$ grows.

\input{gen/sims3d_ablation}
\subsection{$z$-update: Covariance ablations}
We consider three different ways to project onto the terminal covariance constraint \eqref{eq:cov_control_lifted}. A generic conic solver must lift it to an LMI constraint of size $(n_x+W)\times(n_x+W)$ as follows
\begin{equation}
    \|X\|_2\le1 \iff
    \begin{bmatrix} I_{n_x} & X\\ X^\top & I_{\bas N}\end{bmatrix}\in\psd{n_x+\bas N},
    \label{eq:schur_psd}
\end{equation}
with $W:=\bas N$ and project onto $\psd{n_x+W}$ with an eigendecomposition of cost $O((n_x+W)^3)$. However, \socx\ works with spectral-norm ball projections using the $n_x\times n_x$ matrix. We also compare against it by computing the SVD of the full matrix $X$, which has the same asymptotic complexity but a larger constant. In \Cref{tab:sims3d_ablation}, the Gram eigendecomposition takes 0.39--0.42\,ms per call, nearly independent of $N$ over this range; the SVD is 2.4--2.5$\times$ slower, and the Schur-complement PSD projection is 22--216$\times$ slower, with the gap growing with $N$.
\begin{rmrk}
    Batching can be used to increase solve throughput at the expense of added latency by solving multiple instances in parallel, which often increases the individual solve time~\cite{turbompc,jaxipm,cusadi}.
\end{rmrk}

\section{Conclusion}
We presented {\socx}, a GPU-accelerated JAX implementation for solving linear stochastic optimal control with exact representations for linear chance constraints and terminal covariance constraints, with approximate representations for quadratic chance constraints. The proposed solver exploits the structure of the lifted convex representation by using ADMM to split the problem into subproblems that can be solved in closed form. The $\admmx$-update becomes a set of independent linear quadratic tracking problems, and the $z$-update becomes parallel projections. We introduced a spectral ball projection for the terminal covariance constraint to reduce computation time.

The proposed algorithm is validated on a 3D hover linearized quadrotor with different waypoint configurations. Compared to convex solvers for the same problem, we observe speed-ups of up to $98\times$ compared to MOSEK and $2224 \times$ compared to SCS at $10^{-5}$ tolerance. The proposed approach can also solve for longer horizons of up to $N=1120$, compared with $N=320$ for the baselines. Additionally, we presented ablations of the $\admmx$-update and the $z$-update to further isolate components of the observed speed-ups.
\bibliographystyle{IEEEtran}
\bibliography{refs}{}
\end{document}

%% file: gen/lit_table.tex
\DeclareRobustCommand{\cmark}{\checkmark}
\DeclareRobustCommand{\xmark}{$\times$}
\DeclareRobustCommand{\pmark}{\tikz[baseline=-0.6ex]{\draw (0,0) circle (0.6ex); \fill (0,-0.6ex) arc (-90:90:0.6ex) -- cycle;}}
\begin{table}[t]
\centering
\caption{Comparison with related work. CS: covariance steering. Problem class: Gaussian noise (Gau.), exact linear chance constraints (CC), covariance LMI (LMI), convex (Cvx). Solver structure: exploits OCP structure (OCP), time-parallel $O(\log N)$ span ($\log N$), parallel within one problem (Par.), batches across problems (Bat.), GPU. \cmark: yes, \pmark: partial or only some works in the group, \xmark: no, --: not applicable.}
\label{tab:lit_comparison}
\footnotesize
\setlength{\tabcolsep}{1pt}
\begin{tabular}{@{}l cccc ccccc@{}}
\toprule
 & \multicolumn{4}{c}{Problem class} & \multicolumn{5}{c}{Solver structure} \\
\cmidrule(lr){2-5}\cmidrule(lr){6-10}
Work & Gau. & CC & LMI & Cvx & OCP & $\log N$ & Par. & Bat. & GPU \\
\midrule
CS, exact~\cite{okamoto_optimal_2018,Balci2024ConstrainedMV,dokania-lcss2026}
 & \cmark & \cmark & \cmark & \cmark & \xmark & \xmark & \xmark & \xmark & \xmark \\
CS, approx.~\cite{rapakoulias_discrete-time_2023,pilipovsky_computationally_2024}
 & \cmark & \pmark & \cmark & \cmark & \xmark & \xmark & \xmark & \xmark & \xmark \\
Parallel LQR~\cite{sarkka2023,primal-dual-iLQR}
 & \xmark & \xmark & \xmark & \pmark & \cmark & \cmark & \cmark & \pmark & \cmark \\
Batched NLP~\cite{turbompc,cusadi,jaxipm}
 & \xmark & \xmark & \xmark & \xmark & \pmark & \xmark & \pmark & \cmark & \cmark \\
Robust ADMM~\cite{gpusls,cunrto}
 & \xmark & \xmark & \xmark & \pmark & \pmark & \pmark & \cmark & \xmark & \cmark \\
Conic solvers~\cite{CuClarabel,scs-gpu,mosek}
 & -- & \cmark & \pmark & \cmark & \xmark & \xmark & \pmark & \xmark & \pmark \\
\midrule
\textbf{\socx\ (Ours)}
 & \cmark & \cmark & \cmark & \cmark & \cmark & \cmark & \cmark & \cmark & \cmark \\
\bottomrule
\end{tabular}
\end{table}

%% file: algo_diag.tex
\ifdefined\usetikzlibrary
  \usetikzlibrary{positioning, fit, backgrounds, calc, decorations.pathreplacing}
  \makeatletter
  \@ifundefined{tikz@library@arrows.meta@loaded}
    {\tikzset{socx tip/.style={>=to}}}
    {\tikzset{socx tip/.style={>={Latex[length=3.6pt,width=3.6pt]}}}}
  \makeatother
\else
  \errmessage{fig_pipeline.tex: add \noexpand\usepackage{tikz} to your preamble}%
\fi

\providecommand{\socx}{\texttt{socx}}
\providecommand{\bas}[1]{n_e(#1)}
\providecommand{\admmu}{u}
\providecommand{\admmx}{\theta}
\providecommand{\Mop}{\mathcal{M}}
\providecommand{\fmean}{\mu}
\providecommand{\colx}{\varsigma}
\providecommand{\colu}{\upsilon}
\providecommand{\proj}[1]{\Pi_{#1}}
\providecommand{\tgt}{\tau}
\providecommand{\nrho}{n_\rho}
\providecommand{\PRIC}{\textsc{p-ric}}
\providecommand{\PLQT}{\textsc{p-lqt}}
\providecommand{\quaddata}{\mathcal{Q}}
\providecommand{\lindata}{\mathcal{L}}
\providecommand{\birth}[1]{k_{#1}}
\providecommand{\pe}[1]{\textcolor{purple}{#1}}

\begin{figure*}[!t]
\centering
\definecolor{gRed}{HTML}{C0392B}\definecolor{gBlue}{HTML}{1F5FA8}
\scalebox{0.9}{%
\begin{tikzpicture}[
  font=\footnotesize,
  socx tip,
  blk/.style = {draw, rounded corners=2pt, align=center, inner sep=2.5pt},
  grp/.style = {draw, rounded corners=3pt, inner sep=4pt},
  sbox/.style= {draw=orange!70!black!60, dashed, rounded corners=2pt, inner sep=3pt},
  img/.style = {inner sep=0pt},
  hd/.style  = {font=\footnotesize, inner sep=1pt, align=center},
  sub/.style = {font=\scriptsize, text=black!75, inner sep=1pt, align=center},
  eq/.style  = {font=\scriptsize, inner sep=1pt, align=left},
  ar/.style  = {->, semithick, black!65},
]
\node[hd] (xhd) {\textbf{$\admmx$-update}: $\min_{\admmx} J(\admmx)+\tfrac{\rho}{2}\|\Mop\admmx-\tgt\|_2^2$ s.t.\ dynamics\\
   $\Rightarrow$ $1{+}\bas N$ LQT problems};
\node[img, anchor=north west] (lqt) at ($(xhd.south west)+(-4mm,-4mm)$)
  {\includegraphics[width=2.55cm]{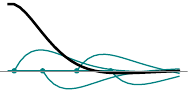}};
\node[sub, anchor=south east] (mlab) at ($(lqt.north west)+(1mm,-2mm)$) {$\mux k$};
\node[sub, text=teal, anchor=north east] (slab) at (lqt.south east) {$\colx^{(c)}_k=0$ for $k<\birth c$};
\node[eq, anchor=west] (calls) at ($(lqt.east)+(1.5mm,0)$)
  {$(\mux{},\muu{})=\PLQT(\quaddata^{\fmean},P^{\fmean},K^{\fmean},\lindata^{\fmean})$\\[4pt]
   \textcolor{teal}{$(\colx^{(c)},\colu^{(c)})=\PLQT(\quaddata^{\colx},P^{\colx},K^{\colx},\lindata^{(c)})$}\\[1pt]
   \textcolor{teal}{\quad batched over $c=0,\dots,\bas N-1$}};
\coordinate (xmid) at ($(mlab.west)!0.5!(calls.east)$);
\node[blk, fill=black!7, draw=black!45, dashed, font=\scriptsize, anchor=north]
  (ric) at ($(xmid|-slab.south)+(0,-2.5mm)$)
  {$(P^{\fmean},K^{\fmean})=\PRIC(\quaddata^{\fmean})$,\;
   $(P^{\colx},K^{\colx})=\PRIC(\quaddata^{\colx})$\\ cached, rerun only when $\rho$ changes};
\begin{scope}[on background layer]
  \node[grp, fill=teal!5, draw=teal!55, fit=(xhd)(mlab)(lqt)(slab)(calls)(ric)] (xbox) {};
\end{scope}

\node[blk, fill=blue!8, draw=blue!45, anchor=south] (prob) at ($(xbox.north)+(0,5mm)$)
  {\Cref{prob:LDLOC}: $\min f(\admmx)+g(z)$ s.t.\ $\Mop\admmx=z$};

\coordinate (zc) at ($(xbox.east|-prob.north)!0.5!(xbox.east|-xbox.south)+(0,-2.5mm)$);
\node[img, anchor=west] (c1) at ($(zc)+(4pt+5mm+4pt+3pt,-3mm)$) {\includegraphics[width=1.35cm]{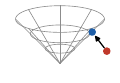}};
\node[img, anchor=south] (c0) at ($(c1.north)+(0,0.5mm)$) {\includegraphics[width=1.35cm]{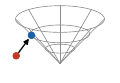}};
\node[img, anchor=north] (c2) at ($(c1.south)+(0,-1.5mm)$) {\includegraphics[width=1.35cm]{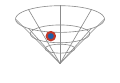}};
\draw[decorate, decoration={brace, amplitude=3pt}, black!55]
  ($(c0.north east)+(1mm,-1mm)$) -- ($(c1.south east)+(1mm,1mm)$) coordinate[midway] (br);
\node[eq, anchor=west] (fout) at ($(br)+(2mm,0)$)
  {$\|s\|_2>|t|$:\\ $(t,s)\mapsto\alpha\,(1,\,s/\|s\|_2)$\\ $\alpha=(t+\|s\|_2)/2$};
\node[eq, anchor=west] (fin) at ($(c2.east)+(3mm,0)$) {$\|s\|_2\le t$:\\ $(t,s)\mapsto(t,s)$};
\node[sub, anchor=south] (chd) at ($(c0.north west)!0.5!(fout.east|-c0.north)+(0,1mm)$)
  {$z_{x,i,k},z_{u,i,k}\in\Ksoc_{\bas k}$\\ all $(i,k)$ in parallel};

\node[eq, anchor=north west] (bform) at ($(fout.east|-zc)+(3pt+2mm+3pt,-6mm)$) {$\sigma_i\leftarrow\min(\sigma_i,1)$};
\node[img, anchor=south] (ball) at ($(bform.north)+(0,1mm)$)
  {\includegraphics[width=1.9cm]{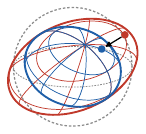}};
\node[sub, anchor=south] (bhd) at ($(ball.north)+(0,1mm)$) {$z_N\in\Bspec$};
\node[sub, anchor=north] (leg) at ($(bform.south)+(0,-1.5mm)$)
  {\textcolor{gRed}{$\bullet$}\,$\hat s+\admmu$ $\to$ \textcolor{gBlue}{$\bullet$}\,$z$};

\node[hd, anchor=north] (zhd) at ($(c0.west|-prob.north)!0.5!(bform.east|-prob.north)+(0,-4pt)$)
  {\textbf{$z$-update}: $z=\proj{\mathcal C}(\hat s+\admmu)$};
\coordinate (it) at ($(zhd.south)+(0,-2pt-3pt)$);      
\coordinate (ib) at ($(xbox.south)+(0,4pt+3pt)$);      
\node[sbox, fit=(chd)(c0)(c2)(fout)(fin)(c0.west|-it)(c0.west|-ib)] (cbox) {};
\node[sbox, fit=(bhd)(ball)(bform)(leg)(ball|-it)(ball|-ib)] (bbox) {};
\begin{scope}[on background layer]
  \node[grp, fill=orange!5, draw=orange!75!black!55, fit=(zhd)(cbox)(bbox)] (zbox) {};
\end{scope}

\node[blk, fill=violet!10, draw=violet!55, font=\scriptsize, anchor=west] (ubox) at ($(zbox.east)+(6mm,9mm)$)
  {\textbf{$\admmu$-update}\\ $\admmu\leftarrow\admmu+\hat s-z$\\[1pt]
   $\hat s=\alpha_{\mathrm{or}}\Mop\admmx$\\ $\;+(1-\alpha_{\mathrm{or}})z^{j}$};
\node[blk, fill=black!4, draw=black!45, font=\scriptsize, below=3.5mm of ubox] (test)
  {\textbf{stop?}\\ $\|\Mop\admmx-z\|_\infty\le\epsilon_\mathrm{pri}$\\ $\rho\|z-z^{j-1}\|_\infty\le\epsilon_\mathrm{dual}$};
\node[blk, fill=blue!8, draw=blue!45, font=\scriptsize, below=3.5mm of test] (out)
  {$\mux{},\muu{},\mvarx{},\mvaru{}$};

\draw[ar] (prob) -- node[sub, right]{split} (prob|-xbox.north);
\draw[ar] (xbox.east|-zbox) -- (zbox.west);
\draw[ar] (zbox.east|-ubox) -- (ubox);
\draw[ar] (ubox) -- (test);
\draw[ar] (test) -- node[sub, right]{yes} (out);
\coordinate (lo) at ($(xbox.south)+(0,-3mm)$);
\draw[ar] (test.east) -- node[sub, above]{no} ++(3mm,0) |- (lo) -- (xbox.south);
\end{tikzpicture}}
\caption{
Steps of ADMM \eqref{eq:admm_def} in \socx. In one iteration, the $\admmx$-update solves mean and stochastic columns' LQT problems in parallel, followed by the $z$-update, which projects every chance-constraint cone and the terminal covariance ball in parallel.}
\label{fig:pipeline}
\end{figure*}

%% file: gen/sims3d_main.tex
\begin{table*}[t]
\centering
\caption{3D quadrotor at tolerance $10^{-5}$: mean $\pm$ max.\ deviation over the three setups of \Cref{fig:traj3d}.   
}
\label{tab:sims3d_main}
\begin{tabular}{rr r@{\,$\pm$\,}l r@{\,$\pm$\,}l cc r@{\,$\pm$\,}l r r@{\,$\pm$\,}l r}
\toprule
 & & \multicolumn{6}{c}{\textbf{\socx\ (Ours)}, GPU} & \multicolumn{3}{c}{MOSEK, CPU} & \multicolumn{3}{c}{SCS+cuDSS, GPU} \\
\cmidrule(lr){3-8}\cmidrule(lr){9-11}\cmidrule(lr){12-14}
$N$ & $\bas N$ & \multicolumn{2}{c}{time (s)} & \multicolumn{2}{c}{iterations} & obj.\ err. & $P_\mathrm{viol}/\epsilon$ & \multicolumn{2}{c}{time (s)} & speed-up & \multicolumn{2}{c}{time (s)} & speed-up \\
\midrule
40 & 252 & 0.38 & 0.02 & 377 & 20 & $2{\times}10^{-7}$ & 1.000 & 2.9 & 0.1 & 7.2--7.9$\times$ & 26 & 11 & 47--104$\times$ \\
80 & 492 & 0.66 & 0.06 & 483 & 53 & $10^{-7}$ & 1.000 & 14.9 & 1.5 & 20--27$\times$ & 179 & 79 & 145--369$\times$ \\
160 & 972 & 2.20 & 0.38 & 750 & 135 & $2{\times}10^{-5}$ & 1.002 & 91.7 & 5.5 & 37--53$\times$ & 1976 & 2077 & 199--\textbf{2224}$\times$ \\
240 & 1{,}452 & 4.33 & 0.30 & 798 & 57 & $10^{-4}$ & 1.002 & 311.7 & 82.9 & 57--\textbf{98}$\times$ & 3848 & 2123 & 386--1342$\times$ \\
320 & 1{,}932 & 7.10 & 0.26 & 808 & 31 & -- & 1.000 & \multicolumn{3}{c}{out of memory (50GB cap)} & 5789 & 1633 & 576--1085$\times$ \\
1120 & 6{,}732 & 144.54 & 55.48 & 1364 & 523 & -- & -- & \multicolumn{3}{c}{out of memory (50GB cap)} & \multicolumn{3}{c}{time out (10,000s cap)} \\
\bottomrule
\end{tabular}
\end{table*}

%% file: gen/sims3d_ablation.tex
\begin{table}[t]
\centering
\caption{Ablations on the helix setup at tolerance $10^{-5}$.}
\label{tab:sims3d_ablation}
\scriptsize
\setlength{\tabcolsep}{3pt}
\begin{tabular}{l rrr rrr}
\toprule
 & \multicolumn{3}{c}{time (ms)} & \multicolumn{3}{c}{scratch memory (MB)} \\
\cmidrule(lr){2-4}\cmidrule(lr){5-7}
Variant & $N{=}80$ & $160$ & $320$ & $N{=}80$ & $160$ & $320$ \\
\midrule
\multicolumn{7}{l}{\emph{$\admmx$-update}: per iteration} \\
\textbf{Parallel LQT (Ours)} & 1.13 & 2.62 & 8.17 & 39 & 124 & 390 \\
Sequential LQT & 4.48 & 9.16 & 20.66 & 8 & 40 & 159 \\
Condensed KKT & 2.46 & 9.54 & 61.19 & 10 & 13 & 43 \\
Per-column loop & 7.78 & 30.63 & 120.72 & 188 & 747 & 3768 \\
\midrule
\multicolumn{4}{l}{\emph{Terminal-LMI projection}: per call} & \multicolumn{3}{@{\hspace{10pt}}l}{\emph{complexity}} \\
\textbf{Gram eigh (Ours)} & 0.39 & 0.42 & 0.41 & \multicolumn{3}{@{\hspace{10pt}}l}{\scalebox{0.8}{$O(n_x^2W+n_x^3)$}} \\
SVD & 0.94 & 1.05 & 0.98 & \multicolumn{3}{@{\hspace{10pt}}l}{\scalebox{0.8}{$O(n_x^2W)$}} \\
Schur PSD & 8.79 & 22.03 & 88.64 & \multicolumn{3}{@{\hspace{10pt}}l}{\scalebox{0.8}{$O((n_x+W)^3)$}} \\
\bottomrule
\end{tabular}
\end{table}